\documentstyle[prd,eqsecnum,aps]{revtex}

\begin{document}

\draft
\title{Are Kaluza-Klein modes enhanced by parametric resonance?}
\author{Shinji Tsujikawa \thanks{electronic
address:shinji@gravity.phys.waseda.ac.jp}}
\address{ Department of Physics, Waseda University,
3-4-1 Ohkubo, Shinjuku, Tokyo 169-8555, Japan\\[.3em]}
\date{\today}
\maketitle
\begin{abstract}
We study parametric amplification of Kaluza-Klein (KK) modes 
in a higher $D$-dimensional generalized
Kaluza-Klein theory, which was originally
considered by Mukohyama in the narrow resonance case.
It was suggested that KK modes can be enhanced by an
oscillation of a scale of compactification by the 
$d$-dimensional sphere $S^d~(d=D-4)$
and by the direct product 
$S^{d_1}\times S^{d_2}~(d_1+d_2=D-4)$.
We extend this past work to the more general case where  
initial values of the scale of compactification and 
the quantum number of the angular momentum 
$l$ of KK modes are not small.
We perform analytic approaches based on the Mathieu equation
as well as numerical calculations, and find that
the expansion of the universe rapidly makes the KK field
deviate from instability bands.
As a result, KK modes are not enhanced sufficiently
in an expanding universe in these two classes of models. 
\end{abstract}


\baselineskip = 24pt

%
\section{Introduction}                            %
Recently, much interest has been focused on higher-dimensional 
theories. The original idea goes back to Kaluza and Klein\cite{KK}, who
considered that the unification of the interactions in Nature
may be realized in the gravitational theory in five dimensions.
At present, there are many higher dimensional theories which are 
the generalizations of the Kaluza-Klein theory,
for example, $D=10$ superstring 
theory\cite{GSW}, and $D=11$ M-theory\cite{M}.
Very recently, the universe of brane models\cite{brane} has
received much attention as a solution to the 
hierarchy problem between the weak and Planck scale. 
In order to hide extra dimensions, we generally assume that 
they are within a very small compact internal space.
Since the typical scale of the internal space is the Planck 
length $l_{\rm pl}\sim 10^{-33}$cm, which is much smaller
than the size of the external space, we can reduce 
higher-dimensional theories to the familiar four-dimensional
gravitational theories.

There are various ways for Kaluza-Klein  reductions to four dimensions.
As was pointed out in Ref.~\cite{mukoh}, it is possible to 
judge whether such reductions are appropriate
from a cosmological point of view.
For example, in the model where the geometry is described by
a direct product of the four-dimensional Minkowski spacetime 
$M^4$ and 1-sphere $S^1$ considered by Kolb 
and Slansky\cite{KS}, particles of Kaluza-Klein (KK) 
modes which are so called {\it pyrgon} are produced. 
Since the energy density of this
massive particle decreases as $\rho_{KK} \sim a^{-3}$, 
this density will dominate over the energy density of the universe 
in the radiation dominant era if pyrgons
are overproduced in the early stage of the universe and
do not decay by entropy productions.
Mukohyama\cite{mukoh} extended this model, and considered
the compactifications both by a  $d$-dimensional 
sphere $S^d$ with $d=D-4$ and by a direct product 
$S^{d_1}\times S^{d_2}$ with $d_1+d_2=D-4$.
In these classes of models, since the reduced effective potential 
in the four-dimensional theory lacks a global
minimum, we need to introduce the Casimir effect as
a one-loop quantum correction to give a stable ground state.
Then a scalar field $\sigma$ which corresponds to 
the scale of the compactification oscillates around 
the minimum of its potential, and is finally trapped
at the ground state $\sigma=0$\cite{stability}.
Since KK modes acquire masses which are
expressed by the oscillating $\sigma$ field, we can expect the 
amplification of KK modes by parametric resonance.
This picture is similar to the scenario of preheating
after inflation where scalar fields coupled to an inflaton 
field are resonantly enhanced in the oscillating stage of 
inflaton\cite{TB,KLS}.
In the case of a chaotic inflation model with a quadratic potential,
particles are most efficiently produced 
in the broad resonance regimes where the amplitude of inflaton
$\phi$  is initially large and the coupling between
$\phi$ and a coupled scalar field $\chi$ is strong\cite{KLS}.
Since the expansion of the universe gradually 
reduces the amplitude of inflaton, this makes 
the $\chi$ field jump over many instability and stability bands.
The $\chi$ field can grow quasiexponentially, overcoming the 
diluting effect by the cosmic expansion.
Finally the field reaches so called narrow 
resonance regimes, after which particle creation terminates.
In the case that the $\chi$ field is initially located
in narrow resonance regimes, it is known that particle creation
is inefficient because the field soon deviates from instability
bands by the cosmic expansion.
As for the excitation of KK modes
in the models of compactifications by $S^d$ and 
$S^{d_1} \times S^{d_2}$, since the 
$\sigma$ field behaves as a massive scalar field 
around the minimum of its potential,
we can apply analytic approaches based on the Mathieu 
equation\cite{mathieu} which have been much investigated
in the context of preheating at the linear stage 
of the system\cite{KLS,structure}.  
However, since the investigation in Ref.~\cite{mukoh}
about the enhancement of KK modes is limited
in narrow resonance regimes in which parametric 
resonance is weak,
we will extend this work in the more general
range of parameters.  
It is of interest whether catastrophic particle production
will occur or not in the case where the value of $\sigma$
is initially large and the quantum number of the 
angular momentum $l$ of KK modes is not small.
If KK modes are strongly enhanced by parametric
resonance, this would result in some important cosmological
implications. They will become good candidates for dark matter
in the case that these excited particles are suitably 
diluted by entropy productions.

This paper is organized as follows.
In the next section, we perform compactifications by a 
$d$-dimensional sphere $S^d$ and by a direct product 
$S^{d_1} \times S^{d_2}$. The Casimir energy is 
introduced to stabilize the scale of the internal space.
In Sec.~III, we study the background evolution of the scale
factor and the $\sigma$ field, and examine the structure of 
resonance for KK modes.
We will investigate whether the enhancement of 
KK modes efficiently occurs or not 
both by analytic approaches and numerical integrations.
We present our conclusion and discussion in the final section.

\section{The model}   

We consider a model in $D=d+4$ dimensions
with a cosmological constant $\bar{\Lambda}$ and 
a single scalar field $\bar{\phi}$, 
\begin{eqnarray}
S=\int d^D x \sqrt{-\bar{g}} \left[ \frac{1}
{2\bar{\kappa}^2} \bar{R}-2\bar{\Lambda}
-\frac12 \bar{g}^{MN}\partial_M \bar{\phi}
\partial_N \bar{\phi} \right],
\label{B1}
\end{eqnarray}
where $\bar{g}_{MN}$ and $\bar{\kappa}^{2}/8\pi \equiv \bar{G}$
are the $D$-dimensional metric and 
gravitational constant, respectively.
$\bar{R}$ is a scalar curvature with respect to $\bar{g}_{MN}$.
The third term of the action $(\ref{B1})$ denotes a Klein-Gordon
action in $D$ dimensions.

We first compactify extra dimensions to a $d$-dimensional 
sphere $S^d$. Then the metric $\bar{g}_{MN}$ can be expressed as
\begin{eqnarray}
ds^2_D=\bar{g}_{MN}dx^M dx^N
=\hat{g}_{\mu\nu}dx^{\mu}dx^{\nu}
+b^2 ds^2_d,
\label{B2}
\end{eqnarray}
where $\hat{g}_{\mu\nu}$ is a four-dimensional metric,
$b$ is a scale of the $d$-dimensional sphere
whose present value is $b_0$, and 
$ds^2_d$ is a line element of the $d$-unit sphere.
We expand the scalar field $\bar{\phi}$ on the sphere as
\begin{eqnarray}
\bar{\phi}=b_0^{-d/2} \sum_{l,m}
\phi_{lm} Y_{lm}^{(d)},
\label{B3}
\end{eqnarray}
where $Y_{lm}^{(d)}$ is the spherical harmonics on the
$d$-sphere with positive integer $l$ and 
a set of $d-1$ numbers $m$.
Then we obtain the reduced action in four dimensions,
\begin{eqnarray}
S=\int d^4 x \sqrt{-\hat{g}} \left(\frac{b}{b_0}\right)^d
 \Biggl[\frac{1}{2\kappa^2} \Biggl\{
\hat{R} &+&
d(d-1) \frac{\partial_{\mu}b \partial_{\nu}b}{b^2}
\hat{g}^{\mu\nu}
+\frac{d(d-1)}{b^2} \Biggr\}
-2V_d^0 \bar{\Lambda}  \nonumber \\
&-& \sum_{l,m} \left\{\frac12 \hat{g}^{\mu\nu}
\partial_{\mu} \phi_{lm} \partial_{\nu} \phi_{lm}
+\frac{l(l+d-1)}{2b^2}\phi_{lm}^2 \right\} \Biggr],
\label{B4}
\end{eqnarray}
where $\kappa^{2}/8\pi=\bar{\kappa}^{2}/(8\pi V_d^0)$
is Newton's gravitational constant with $V_d^0$
the present volume of the internal space.
$\hat{R}$ is a scalar curvature related 
with $\hat{g}_{\mu\nu}$.
Since this does not take the ordinary form of the Einstein-Hilbert
action because of a time dependent factor $(b/b_0)^d$,
we perform a conformal transformation as follows
\begin{eqnarray}
\hat{g}_{\mu\nu}=\exp\left(-d\frac{\sigma}{\sigma_0}
\right) g_{\mu\nu},
\label{B5}
\end{eqnarray}
where $\sigma$ is the so-called dilaton field
which is defined by  
\begin{eqnarray}
\sigma &=& \sigma_0 {\rm ln} \left( \frac{b}{b_0} \right), \\
\sigma_0 &=& \left[ \frac{d(d+2)}{2\kappa^2} \right]^{1/2}.
\label{B7}
\end{eqnarray}
Then the four-dimensional action in the Einstein frame
can be described as
\begin{eqnarray}
S=\int d^4 x \sqrt{-g} \left[ \frac{1}{2\kappa^2}R
-\frac12 g^{\mu\nu} \partial_{\mu} \sigma 
\partial_{\nu} \sigma -U_1(\sigma)
-\frac12 \sum_{l,m} ( g^{\mu\nu} \partial_{\mu} 
\phi_{lm} \partial_{\nu} \phi_{lm} +
M_l^2(\sigma)\phi_{lm}^2)\right],
\label{B8}
\end{eqnarray}
where $R$ is a scalar curvature with respect to 
$g_{\mu\nu}$. 
$U_1(\sigma)$ is a potential of the $\sigma$ field 
which we will give a specific form below, and the mass
$M_l(\sigma)$ of the $\phi_{lm}$ field is expressed as
\begin{eqnarray}
M_l^2(\sigma)=\frac{l(l+d-1)}{b_0^2}
e^{-(d+2)\sigma/\sigma_0}.
\label{B9}
\end{eqnarray}

As was mentioned in Ref.~\cite{mukoh}, the potential 
$U_1(\sigma)$ does not have a local minimum 
if we do not introduce
the Casimir effect as a one-loop quantum correction. 
Taking into account this effect, $U_1(\sigma)$ can be 
written in the form
\begin{eqnarray}
U_1(\sigma)=\alpha \left[ \frac{2}{d+2}
e^{-2(d+2)\sigma/\sigma_0}+e^{-d\sigma/\sigma_0}
-\frac{d+4}{d+2} e^{-(d+2)\sigma/\sigma_0} \right],
\label{B10}
\end{eqnarray}
with
\begin{eqnarray}
\alpha=\frac{(d-1)\sigma_0^2}{(d+4)b_0^2}.
\label{B11}
\end{eqnarray}
The first, second, and third terms in $(\ref{B10})$ 
appear  due to the Casimir energy, the cosmological 
constant, and the curvature of the internal space, respectively.
The potential $U_1(\sigma)$ 
acquires a local minimum at $\sigma=0$ for the case of $d \ge 2$
because of the presence of the first term in $(\ref{B10})$.
We depict $U_1(\sigma)$ in Fig.~1 for the case of $d=2$.
It has a local maximum at $\sigma_* (>0)$, which depends on
the extra dimension $d$.  $\sigma_*/\sigma_0$ decreases
with the increase of $d$. For example, 
$\sigma_*/\sigma_0=0.50$ for $d=2$, 
and $\sigma_*/\sigma_0=0.22$ for $d=6$. 
Since $\sigma_0$ becomes larger with the increase of $d$ 
[See Eq.~$(\ref{B7})$], $\sigma_*$ itself does not necessarily
decrease with the increase of $d$.
For example, $\sigma_*=0.20m_{\rm pl}$ for $d=2$, 
and $\sigma_*=0.21m_{\rm pl}$ for $d=6$. 
In order to result in the final state $\sigma=0$ which 
corresponds to the present value $b=b_0$, initial values
of $\sigma$ are required 
to be $0<\sigma_i<\sigma_*$ (we assume $\sigma_i>0$),
where the subscript $i$ denotes the initial value.
Then $\sigma$ evolves toward the minimum of its 
potential, and begins to oscillate around $\sigma=0$.
In this oscillating stage, we are concerned with whether 
KK modes are effectively enhanced or not
by parametric resonance. 

In the case that the compactification is done on 
a direct product $S^{d_1}\times S^{d_2}$ of the 
$d_1$-dimensional sphere $S^{d_1}$ and
$d_2$-dimensional sphere $S^{d_2}$ with 
$d_1+d_2=D-4$, the final four-dimensional action 
and the potential $U_1(\sigma)$ take different forms
from the $S^d$ case.
The procedure is the same as in the $S^d$ case.
First, we consider the $D$-dimensional metric 
\begin{eqnarray}
ds^2_D=\bar{g}_{MN}dx^M dx^N=
\hat{g}_{\mu\nu}dx^{\mu}dx^{\nu}
+b_1^2 ds^2_{d_1}+b_2^2 ds^2_{d_2},
\label{B12}
\end{eqnarray}
where $b_1$ and $b_2$ are scales of the sphere $S^{d_1}$
and $S^{d_2}$, 
$ds^2_{d_1}$ and $ds^2_{d_2}$ are line elements of 
the $d_1$-unit sphere and $d_2$-unit sphere, respectively.

Expanding the $\bar{\phi}$ field in the $D$-dimensional
action $(\ref{B1})$ as 
\begin{eqnarray}
\bar{\phi}=b_{10}^{-d_1/2} b_{20}^{-d_2/2}
\sum_{l_1,l_2,m_1,m_2}
\phi_{l_1 l_2 m_1 m_2} Y_{l_1 m_1}^{(d_1)}
Y_{l_2 m_2}^{(d_2)},
\label{B13}
\end{eqnarray}
where $b_{10}$ and $b_{20}$ are present values of $b_1$ and
$b_2$, $Y_{l_1 m_1}^{(d_1)}$ and $Y_{l_2 m_2}^{(d_2)}$
are the spherical harmonics on the $d_1$-sphere and $d_2$-sphere,
respectively, we obtain the four-dimensional action similar to
the action (2.4) in the $S^d$ case
(see Ref.~\cite{mukoh} for details).
Next, we perform a conformal transformation 
\begin{eqnarray}
\hat{g}_{\mu\nu}=\exp\left[-\left(\frac{d_1\sigma_1}
{\sigma_{10}}+\frac{d_2\sigma_2}{\sigma_{20}} \right)
\right] g_{\mu\nu},
\label{B14}
\end{eqnarray}
where $\sigma_i (i=1,2)$ and $\sigma_{i0}$
are defined by
\begin{eqnarray}
\sigma_i &=& \sigma_{i0} {\rm ln} \left( \frac{b_i}{b_{i0}} \right), \\
\sigma_{i0}&=& \left[ \frac{d_i(d_i+2)}{2\kappa^2} \right]^{1/2}.
\label{B16}
\end{eqnarray}
Taking into account the Casimir effect as in the $S^d$ case
and imposing the condition 
\begin{eqnarray}
\frac{\sigma_1}{\sigma_{10}}=\frac{\sigma_2}{\sigma_{20}}
\equiv \frac{\sigma}{\sigma_0},
\label{B17}
\end{eqnarray}
with
\begin{eqnarray}
\sigma_0=\sqrt{\frac{(d_1+d_2)(d_1+d_2+2)}{2\kappa^2}},
\label{B18}
\end{eqnarray}
the final expression of the four-dimensional action yields
\begin{eqnarray}
S=\int d^4 x &\sqrt{-g}& \Biggl[ \frac{1}{2\kappa^2}R
- \frac12 g^{\mu\nu} \partial_{\mu} \sigma 
\partial_{\nu} \sigma -U_1(\sigma)  \nonumber \\
&-& 
\frac12 \sum_{l_1,l_2,m_1,m_2} ( g^{\mu\nu} \partial_{\mu} 
\phi_{l_1l_2m_1m_2} \partial_{\nu} 
\phi_{l_1l_2m_1m_2} + 
M_{l_1l_2}^2(\sigma)\phi_{l_1l_2m_1m_2}^2) \Biggr],
\label{B19}
\end{eqnarray}
where the potential of the $\sigma$ field is 
\begin{eqnarray}
U_1(\sigma)=\alpha \left[ \frac{2}{d_1+d_2+2}
e^{-2(d_1+d_2+2)\sigma/\sigma_0}+
e^{-(d_1+d_2)\sigma/\sigma_0}
-\frac{d_1+d_2+4}{d_1+d_2+2} 
e^{-(d_1+d_2+2)\sigma/\sigma_0} \right],
\label{B20}
\end{eqnarray}
with
\begin{eqnarray}
\alpha =\frac{d_1+d_2+2}{2(d_1+d_2+4)\kappa^2}
\left[ \frac{d_1(d_1-1)}{b_{10}^2}+
\frac{d_2(d_2-1)}{b_{20}^2} \right].
\label{B21}
\end{eqnarray}
The mass of $M_{l_1 l_2}(\sigma)$ of the 
$\phi_{l_1l_2m_1m_2}$ field is given by
\begin{eqnarray}
M_{l_1 l_2}^2(\sigma)=\left[
\frac{l_1(l_1+d_1-1)}{b_{10}^2}+
\frac{l_2(l_2+d_2-1)}{b_{20}^2}\right]
e^{-(d_1+d_2+2)\sigma/\sigma_0}.
\label{B22}
\end{eqnarray}
The terms in the square bracket of Eq.~$(\ref{B20})$ correspond to
changing $d$ in that of Eq.~$(\ref{B10})$ to $d_1+d_2$.
Hence the shape of the potential $U_1(\sigma)$ 
in the $S^{d_1} \times S^{d_2}$ case is the same as
in the $S^d$ case. 
$U_1(\sigma)$ has a local minimum at $\sigma=0$
when either of $d_1$ and $d_2$ is greater than 1.
In the next section, the dynamics of $\sigma$ and $\phi_{lm}$ fields
are studied in two classes of models 
of $(\ref{B8})$ and $(\ref{B19})$.

\section{The dynamics of the system}   

In this section, we investigate the excitation of KK modes
due to the oscillating scalar field $\sigma$ by parametric
resonance.
We assume that the four-dimensional spacetime 
and the $\sigma$ field are
spatially homogeneous, and adopt the flat
Friedmann-Robertson-Walker metric as 
the background spacetime
\begin{eqnarray}
ds_4^2=g_{\mu\nu}dx^{\mu}dx^{\nu}=
-dt^2 + a^2(t) d {\bf x^2},
\label{C1}
\end{eqnarray}
where $t$ is the cosmic time and $a(t)$ is the scale factor.
The background equations for $a$ and $\sigma$ can be 
expressed by the action $(\ref{B8})$ or $(\ref{B19})$ as
\begin{eqnarray}
\left(\frac{\dot{a}}{a}\right)^2
= \frac{\kappa^2}{3}
\left[\frac12 \dot{\sigma}^2+U_1(\sigma) \right],
\label{C2}
\end{eqnarray}
\begin{eqnarray}
\ddot{\sigma}+3\frac{\dot{a}}{a}
\dot{\sigma}+U_1'(\sigma)=0,
\label{C3}
\end{eqnarray}
where we have neglected the contributions from
KK modes. However, in the case that 
KK modes are enhanced significantly,
this affects the evolution of background equations.
We briefly comment on this back reaction issue
at the end of this section. 

As we have already mentioned, the initial value of $\sigma$ is
required to be smaller than $\sigma_*$ in order not to lead to 
an expansion of the internal space.
Let us consider the case where $\sigma$ is initially located close 
to $\sigma_*$ with $\sigma_i<\sigma_*$.
In this case, defining a dimensionless parameter 
\begin{eqnarray}
\epsilon =
\cases{ (d+2)\frac{\sigma}{\sigma_0}
 & ($S^d$~case), \cr
 (d_1+d_2+2)  \frac{\sigma}{\sigma_0}& 
($S^{d1} \times S^{d2}$~case), \cr
}
\label{C40}
\end{eqnarray}
it usually exceeds of the order of unity at the beginning
with $d \ge 2$ or $d_1+d_2 \ge 2$.
For example, in the $S^d$ case, $\epsilon(t_i)=2.0$ when
$d=2$, and $\epsilon(t_i)=1.8$ when $d=6$.
This is the situation which was not considered in 
Ref.~\cite{mukoh}.
After a short stage of rolling down, 
the $\sigma$ field begins to oscillate around the minimum of 
its potential at $\sigma=0$.
After that, the $\sigma$ field
behaves as a massive scalar field whose mass is given by
\begin{eqnarray}
m_{\sigma} \equiv \sqrt{U_1''(0)}=
\cases{
\frac{\sqrt{2(d-1)}}{b_0} & ($S^d$~case), \cr
\sqrt{\frac{2}{d_1+d_2}
\left[\frac{d_1(d_1-1)}{b_{10}^2}+
\frac{d_2(d_2-1)}{b_{20}^2} \right] } & 
($S^{d1} \times S^{d2}$~case). \cr
}
\label{C4}
\end{eqnarray}
The evolution of this field is the same as a
massive inflaton field $\phi$ with a chaotic potential
$V(\phi)=m_{\phi}^2\phi^2/2$
in the reheating phase.
Making use of the time-averaged relation
\begin{eqnarray}
\frac12 \langle \dot{\sigma}^2 \rangle
=\langle U_1(\sigma) \rangle,
\label{C5}
\end{eqnarray}
we easily find from Eqs.~$(\ref{C2})$ and $(\ref{C3})$
that the evolution of $a$ and $\sigma$ 
can be approximately written as
\begin{eqnarray}
a = \left(\frac{t}{t_{\rm co}}\right)^{2/3},
\label{C6}
\end{eqnarray}
\begin{eqnarray}
\sigma(t)/\sigma_0=\tilde{\sigma}(t) \cos m_{\sigma} t,
\label{C7}
\end{eqnarray}
where $t_{\rm co}$ denotes the time when the coherent 
oscillation of $\sigma$ starts, and 
$\tilde{\sigma}(t)$ is the dimensionless amplitude
of the $\sigma$ field which decreases as 
$\tilde{\sigma}(t) \sim 1/t$ by the cosmic expansion.
Eq.~$(\ref{C6})$ shows that the universe evolves as
matter dominant when the $\sigma$ field oscillates coherently.
Hence the energy density of $\sigma$ decreases as
$\rho_{\sigma} \sim 1/t^3$.

Let us study the excitation of KK modes
during the oscillating stage of the $\sigma$ field. 
For each compactification of the $S^d$ and $S^{d_1} \times
S^{d_2}$ cases, 
we expand $\phi_{lm}$ ($S^d$ case)
and $\phi_{l_1 l_2 m_1 m_2}$ ($S^{d_1}\times
S^{d_2}$ case) fields by the Fourier component as
\begin{eqnarray}
\phi_{lm}=\frac{1}{(2\pi)^{3/2}} \int \left(a_k 
\phi_{lm}^{\bf k}(t) e^{-i {\bf k} \cdot {\bf x}}
+a_k^{\dagger} \phi_{lm}^{\bf k *}(t) 
 e^{i {\bf k} \cdot {\bf x}} \right) d^3{\bf k},
\label{C8}
\end{eqnarray}
\begin{eqnarray}
\phi_{l_1l_2m_1m_2}=\frac{1}{(2\pi)^{3/2}}
\int \left(a_k \phi_{l_1l_2m_1m_2}^{\bf k}(t) 
e^{-i {\bf k} \cdot {\bf x}}
+a_k^{\dagger} \phi_{l_1l_2m_1m_2}^{\bf k *}(t) 
e^{i {\bf k} \cdot {\bf x}} \right) d^3{\bf k},
\label{C50}
\end{eqnarray}
where $a_k$ and $a_k^{\dagger}$ are the annihilation and
creation operators, respectively.
Then we find that the temporary parts 
$\phi_{lm}^{\bf k}(t)$ and 
$\phi_{l_1 l_2 m_1 m_2}^{\bf k}(t)$ obey
the following equations of motion:  
\begin{eqnarray}
\ddot{\phi}_{lm}^{\bf k}+
 3\frac{\dot{a}}{a} \dot{\phi}_{lm}
 ^{\bf k}+\left[ \frac{k^2}{a^2}+\frac{l(l+d-1)}{b_0^2}
e^{-(d+2)\sigma/\sigma_0}
 \right]
 \phi_{lm}^{\bf k}=0~~~~~~(S^d~{\rm case}),
\label{C9}
\end{eqnarray}
\begin{eqnarray}
& &\ddot{\phi}_{l_1l_2m_1m_2}^{\bf k} +
 3\frac{\dot{a}}{a} \dot{\phi}_{l_1l_2m_1m_2}
 ^{\bf k}+ \nonumber \\
 & & \left[ \frac{k^2}{a^2}+ 
 \left\{\frac{l_1(l_1+d_1-1)}{b_{10}^2}+
 \frac{l_2(l_2+d_2-1)}{b_{20}^2} \right\}
e^{-(d_1+d_2+2)\sigma/\sigma_0}
 \right] \phi_{l_1l_2m_1m_2}^{\bf k}=0
 ~~~~~~(S^{d_1} \times
S^{d_2}~{\rm case}). \nonumber \\
\label{C51}
\end{eqnarray}
Hereafter, we express both fields $\phi_{lm}^{\bf k}$
and $\phi_{l_1 l_2 m_1 m_2}^{\bf k}$ as $\phi_k$ except 
the case that distinctions of both fields are required.
Defining a new scalar field $\varphi_k
=a^{3/2} \phi_k$, Eqs.~$(\ref{C9})$ and (3.12)
are written as 
\begin{eqnarray}
\ddot{\varphi}_k+
 \omega_k^2 \varphi_k=0,
\label{C10}
\end{eqnarray}
where for the $S^d$ case,
\begin{eqnarray}
\omega_k^2 \equiv \frac{k^2}{a^2}
+\frac{l(l+d-1)}{b_0^2}
e^{-(d+2)\sigma/\sigma_0}
-\frac34 \left(\frac{2\ddot{a}}{a}
+\frac{\dot{a}^2}{a^2} \right),
\label{C11}
\end{eqnarray}
and for the $S^{d_1}\times S^{d_2}$ case,
\begin{eqnarray}
\omega_k^2 \equiv \frac{k^2}{a^2}+
\left\{\frac{l_1(l_1+d_1-1)}{b_{10}^2}+
 \frac{l_2(l_2+d_2-1)}{b_{20}^2} \right\}
e^{-(d_1+d_2+2)\sigma/\sigma_0}
-\frac34 \left(\frac{2\ddot{a}}{a}
+\frac{\dot{a}^2}{a^2} \right).
\label{C52}
\end{eqnarray}
As for the initial conditions of the $\varphi_k$
field, we choose the 
state that corresponds to the conformal vacuum as
\begin{eqnarray}
\varphi_k(t_i)=\frac{1}{\sqrt{2\omega_k (t_i)}},~~~
\dot{\varphi}_k(t_i)=-i \omega_k(t_i) \varphi_k(t_i).
\label{B35}
\end{eqnarray}
The second terms in
Eqs.~$(\ref{C11})$ and $(\ref{C52})$
are resonance terms which would lead
to the amplification of KK modes.
When $\epsilon$ defined by Eq.~(3.4) is greater than of the order of unity, 
resonance terms are suppressed and do not play relevant roles for
the enhancement of the $\varphi_k$ field.   
However, $\epsilon$ decreases under 
unity and $\sigma$ begins to oscillate coherently,
we can expect the parametric resonance phenomenon.
In this case, Eq.~$(\ref{C10})$ is reduced to the Mathieu
equation,
\begin{eqnarray}
\frac{d^2}{dz^2} \varphi_k +\left(A_k -2q \cos 2z \right) 
\varphi_k=0,
\label{C12}
\end{eqnarray}
where for the $S^d$ case,
\begin{eqnarray}
A_k=\frac{2l(l+d-1)}{d-1}+
4\frac{(k/m_{\sigma})^2}{a^2},
\label{C13}
\end{eqnarray}
\begin{eqnarray}
q=\frac{l(l+d-1)}{d-1} (d+2)\tilde{\sigma}(t),
\label{C14}
\end{eqnarray}
and for the $S^{d_1}\times S^{d_2}$ case,
\begin{eqnarray}
A_k=\frac{2(d_1+d_2) \{l_1(l_1+d_1-1)
+l_2(l_2+d_2-1)(b_{10}/b_{20})^2 \}} 
{d_1(d_1-1)+d_2(d_2-1)(b_{10}/b_{20})^2}
+4\frac{(k/m_{\sigma})^2}{a^2},
\label{C15}
\end{eqnarray}
\begin{eqnarray}
q=\frac{(d_1+d_2) \{l_1(l_1+d_1-1)
+l_2(l_2+d_2-1)(b_{10}/b_{20})^2 \}} 
{d_1(d_1-1)+d_2(d_2-1)(b_{10}/b_{20})^2}
(d_1+d_2+2) \tilde{\sigma}(t).
\label{C16}
\end{eqnarray}
$z$ is defined as $z=mt$.
We have neglected the contribution of the third term in 
Eq.~$(\ref{C11})$ and $(\ref{C52})$ 
since this term is not important in the 
oscillating stage of the $\sigma$ field\cite{KLS}.

We can analyze the excitation of KK modes 
by making use of the stability and instability 
chart of the Mathieu equation\cite{mathieu}.
Parametric resonance occurs when KK modes 
stay in the instability regions sketched in Fig.~2.
These regions are typically described by
narrow resonance regimes of 
$q~\mbox{\raisebox{-1.ex}{$\stackrel
     {\textstyle<}{\textstyle \sim}$}}~1$ and
broad resonance regimes of $q \gg 1$. 
As is found in Fig.~2,  the instability bands are broader
for larger values of $q$ as long as 
$A_k$ is not so large.
Generally, particle creation in narrow resonance regimes
is not so efficient as in the case of 
broad resonance regimes\cite{KLS}.
Since the past work about the excitation of KK 
modes\cite{mukoh} is limited in  narrow resonance
regimes $q~\mbox{\raisebox{-1.ex}{$\stackrel
     {\textstyle<}{\textstyle \sim}$}}~1$,
we extend this work in the more
general case with $q \gg 1$.
However, note that the results obtained in the context of 
preheating with a quadratic potential\cite{KLS} are not 
directly applied to the present case, because the relation of
$A_k$ and $q$ is different. 

First, let us first examine the $S^d$ case. 
As is found in Eq.~$(\ref{C13})$, $A_k$ approaches 
the constant value 
\begin{eqnarray}
B \equiv \frac{2l(l+d-1)}{(d-1)},
\label{C35}
\end{eqnarray}
with the passage of time for fixed values of $d$ and $l$.
Since $B$ is greater than unity for the positive integer
$d$ and $l$ with $d \ge 2$,
parametric resonance does not occur
in the first resonance band around $A_k=1$
as was pointed out  in Ref.~\cite{mukoh}. 
For example, when $d=2$ and $l=1$, $B=4$.
The initial value of $q$ ($=q_{\rm co}$) at which the analysis
based on the Mathieu equation becomes valid
depends on the value of 
$\sigma$ when the $\sigma$ field begins to oscillate
coherently. Numerical calculations show that 
this corresponds to $\epsilon (t_{\rm co}) \approx 0.4$,
after which the evolution of the $\sigma$ field 
can be described as Eq.~$(\ref{C7})$.
This means that the $(d+2) \tilde{\sigma}(t)$ term in 
Eq.~$(\ref{C14})$ and the $(d_1+d_2+2) \tilde{\sigma}(t)$ 
term in Eq.~$(\ref{C16})$ are restricted to be smaller than
unity. In order to realize the situation $q \gg 1$, we need to
choose larger values of $l$, or $l_1$ and $l_2$.

Let us investigate two cases of 
$q_{\rm co}~\mbox{\raisebox{-1.ex}{$\stackrel
     {\textstyle<}{\textstyle \sim}$}}~1$ 
and $q_ {\rm co} \gg 1$.
First, we consider the case of 
$q_{\rm co}~\mbox{\raisebox{-1.ex}{$\stackrel
     {\textstyle<}{\textstyle \sim}$}}~1$ 
with the extra dimension
$d=2$. When $l=1$, since
$A_k=4+4(k/m_{\sigma}a)^2$ and 
$q_{\rm co}=8\tilde{\sigma}(t_{\rm co})\approx 0.8$ by 
Eqs.~$(\ref{C13})$ and $(\ref{C14})$,
the $\varphi_k$ field is initially located in the second 
instability band around $A_k=4$ for low momentum modes.
However, the value of 
$q$ becomes smaller with the decrease of the amplitude 
$\tilde{\sigma}(t)$ as $\tilde{\sigma}(t) \sim1/t$ 
by the cosmic expansion.
In Fig.~3, we depict the evolution of the $\sigma$ field
in this case. We choose $\sigma(t_i)/\sigma_0=0.4$ as 
the initial value of the $\sigma$ field.  
After the first stage of rolling down, 
$\sigma$ begins to oscillate with the initial amplitude
$\tilde{\sigma}(t_{\rm co}) \approx 0.1$.
The coherent oscillation starts by the same amplitude
as long as the initial value of $\sigma$ is located in the region of 
$0.1~\mbox{\raisebox{-1.ex}{$\stackrel
     {\textstyle<}{\textstyle \sim}$}}
     ~\sigma(t_i)/\sigma_0~\mbox{\raisebox{-1.ex}{$\stackrel
     {\textstyle<}{\textstyle \sim}$}}~\sigma_*/\sigma_0
\approx 0.5$.
After several oscillations, the amplitude drops down 
$\tilde{\sigma}(t)~\mbox{\raisebox{-1.ex}{$\stackrel
     {\textstyle<}{\textstyle \sim}$}}~0.02$, which 
 corresponds to 
$q~\mbox{\raisebox{-1.ex}{$\stackrel
     {\textstyle<}{\textstyle \sim}$}}~0.1$. 
Although the $\varphi_k$ field 
stays in the instability band 
since $A_k$ approaches $A_k \to 4$, the growth rate 
$\mu_k$ of the $\varphi_k$ field becomes 
smaller with the decrease of $q$.
In the $j$-th resonance band with $A_k=j^2$,
$q \ll 2j^{3/2}$, and $j \ge 2$,
$\mu_k$ is expressed as\cite{mathieu}
\begin{eqnarray}
\mu_k=\frac{\sin2\delta}{2j [2^{j-1}(j-1)]^2}q^j,
\label{C17}
\end{eqnarray}
where $\delta$ changes in the interval $[0,\pi/2]$.
In the second instability band with $q\le q_{\rm co} \approx 0.8$,
$\mu_k \propto q^2$. 
The growth rate $\mu_k$ is initially of the order 0.01, 
and after that $\mu_k$ rapidly decreases.
The growth rate of KK modes 
$\phi_k=a^{-3/2}\varphi_k$ is approximately expressed as
\begin{eqnarray}
\frac{d\phi_k}{dz} \approx a^{-3/2}
\left(\mu_k -\frac32 \frac{H}{m}\right) \varphi_k,
\label{C18}
\end{eqnarray}
where we have used the relation $\varphi_k \sim e^{\mu_k z}$.
Since the Hubble expansion rate at $t=t_{\rm co}$ is 
$H/m \approx 0.28$, which exceeds the growth rate of $\varphi_k$, 
$\phi_k$ decreases at the beginning.
Moreover, since $\mu_k$ and $H$ decrease as
 $\mu_k \propto q^2 \propto 1/t^2$ and $H \propto 1/t$,
the increase of $\phi_k$ can not be expected 
in the second instability band of the 
narrow resonance regime 
$q~\mbox{\raisebox{-1.ex}{$\stackrel
     {\textstyle<}{\textstyle \sim}$}}~1$.
We have numerically confirmed that $\phi_k$ 
does not increase in the case of $d=2$ and $l=1$ 
(see Fig.~4).
If there is a set of $d$ and $l$ which satisfies
$B=j^2$ with positive integer $j \ge 3$, the  
narrow instability bands exist around the regions 
of $A_k=j^2$.
However, parametric resonance is much more 
inefficient than in the second resonance case, 
since the growth rate $\mu_k$ decreases 
with the increase of $j$ for 
$q~\mbox{\raisebox{-1.ex}{$\stackrel
     {\textstyle<}{\textstyle \sim}$}}~1$
as is found by Eq.~$(\ref{C17})$.
Even if $d$ and $l$ are changed, 
the first resonance does not occur
because $B$ defined in Eq.~$(\ref{C35})$
is always greater than unity.
In the narrow resonance of 
$q~\mbox{\raisebox{-1.ex}{$\stackrel
     {\textstyle<}{\textstyle \sim}$}}~1$
with the $S^d$ compactification,
we can conclude that KK modes are not enhanced sufficiently
in any values of the extra dimension $d$ and the
positive integer $l$.

Consider the case of $q \gg 1$
with the $S^d$ compactification. 
As we have already mentioned, larger values of $l$ are required
to lead to the condition $q \gg 1$.
In order to judge whether parametric 
resonance is efficient or not,
we should know the relation between $A_k$ and $q$.
In the $S^d$ case, we obtain the following relation
from Eqs.~$(\ref{C13})$ and $(\ref{C14})$,
\begin{eqnarray}
A_k=\frac{2}{(d+2)\tilde{\sigma}(t)} q+
4\frac{(k/m_{\sigma})^2}{a^2}.
\label{C19}
\end{eqnarray}
The efficiency of resonance strongly depends on 
the value of $(d+2)\tilde{\sigma}(t)$ in Eq.~$(\ref{C19})$.
We have numerically confirmed that this value is approximately
$(d+2)\tilde{\sigma}(t_{\rm co}) \approx 0.4$
when the $\sigma$ field  begins to oscillate coherently
for any values of $d$. 
This corresponds the relation of 
$A_k \approx 
5q_{\rm co}+4(k/m_{\sigma}a)^2$
at the beginning of the oscillating stage.
Although there are many instability bands in the 
regions of $q \gg 1$ which are generally called 
broad resonance bands, these bands become rare 
in the regions of $A_k~\mbox{\raisebox{-1.ex}{$\stackrel
     {\textstyle>}{\textstyle\sim}$}}~5q$ (see Fig.~2).
Moreover, the tangent of the line 
$A_k=2q/\{(d+2)\tilde{\sigma}(t)\}$ gets
larger with the passage of time because of the 
decrease of the amplitude $\tilde{\sigma}(t)$
as $\tilde{\sigma}(t) \sim 1/t$,
which means that parametric resonance becomes inefficient.
Even in the case that the $\varphi_k$ field is initially in 
an instability band, the expansion of the universe makes 
the field deviate from the instability band.
In the model of a massive inflaton with a coupled scalar
field $\chi$ in preheating after inflation, 
the $\chi$ field moves on the path of 
$A_k=2q+k^2/(m_{\phi}^2a^2)$ with the decrease of 
$q$ as $q \sim 1/t^2$.
In this case, there are many instability bands as well as
stability bands in the regions
of $A_k \ge 2q$, the $\chi$ field {\it stochastically} increases
when passing through instability bands.
Since the tangent of the $A_k$-$q$ path does not change,
$\chi$ particles can be efficiently produced overcoming the 
diluting effect by the cosmic expansion\cite{KLS}.
In the present model, the excitation of KK modes 
is very weak because the regions of 
$A_k \ge 2q/\{(d+2)\tilde{\sigma}(t)\}$
are mostly occupied by stability bands.
Although the $\varphi_k$ field passes
through instability bands,
these are very few in the regions of 
$A_k \ge 2q/\{(d+2)\tilde{\sigma}(t)\}$, and parametric 
resonance is not efficient.
Even the low momentum modes close to the line 
$A_k=2q/\{(d+2)\tilde{\sigma}(t)\}$ are not 
enhanced sufficiently.
As time passes, both $A_k$ and $q$ decrease
by the cosmic expansion, and
finally approaches $A_k=B=2l(l+d-1)/(d-1)$ and $q=0$.
In Fig.~5, we show the evolution of the KK mode
$\phi_k$ with $d=2$, $l=100$, and $k=0$.
In this case, $q_{\rm co} \approx 8000$. We find that
the KK mode does not grow as in the case of 
$q~\mbox{\raisebox{-1.ex}{$\stackrel
     {\textstyle<}{\textstyle \sim}$}}~1$. 
Namely, the growth rate $\mu_k$ of the $\varphi_k$ field
does not surpass the Hubble expansion rate
even in the case of $q \gg 1$.
As we have already mentioned, 
the initial relation of $A_k$ and $q$ 
hardly depends on the values of $\sigma(t_{\rm co})$ and
$d$, which means that parametric resonance is also
inefficient even if we choose larger values of 
$\sigma(t_i)$ and $d$.

Next, let us consider the compactification by $S^{d_1}
\times S^{d_2}$. In this case, since the value 
\begin{eqnarray}
B \equiv
\frac{2(d_1+d_2) \{l_1(l_1+d_1-1)
+l_2(l_2+d_2-1)(b_{10}/b_{20})^2 \}}
{d_1(d_1-1)+d_2(d_2-1)(b_{10}/b_{20})^2}
\label{C36}
\end{eqnarray}
is greater than unity for positive integers 
$l_1, l_2$ and $d_1, d_2$ with $d_1 \ge d_2$
and $d_1+d_2=D-4$, the first 
resonance does not occur as in the 
case of $S^d$\cite{comment}.
When $B$ is written by a positive integer $j (\ge 2)$
as $B=j^2$ and the value of $q_{\rm co}$ is 
smaller than unity,
the $\varphi_k$ field is located in an instability band 
at the initial stage. However, since $q$ decreases as
$q \sim 1/t$ by 
the cosmic expansion and the growth rate of KK
modes is very small for $j \ge 2$, the excitation of
KK modes is inefficient in narrow 
resonance regimes $q~\mbox{\raisebox{-1.ex}{$\stackrel
     {\textstyle<}{\textstyle \sim}$}}~1$.
In the case of $q \gg 1$, since the relation of $A_k$ and 
$q$ is written by Eqs.~$(\ref{C15})$ and $(\ref{C16})$
as 
\begin{eqnarray}
A_k=\frac{2}{(d_1+d_2+2)\tilde{\sigma}(t)} q+
4\frac{(k/m_{\sigma})^2}{a^2},
\label{C20}
\end{eqnarray}
the discussions in the $S^d$ case can be applied by changing 
$d_1+d_2$ to $d$.
We have numerically confirmed that the $\sigma$ field 
begins to oscillate coherently at 
$(d_1+d_2+2) \tilde{\sigma}(t_{\rm co})\approx 0.4$
for any values of $\tilde{\sigma}(t_i)$, and $d_1$, $d_2$.
Moreover, since $\tilde{\sigma}(t)$ decreases as
$\tilde{\sigma}(t) \sim 1/t$, parametric resonance is also
ineffective as in the $S^d$ case.

We conclude that catastrophic particle creation does
not occur in two classes of models by $S^d$ and
$S^{d_1} \times S^{d_2}$ compactifications
in any values of parameters. 
As a result, the back reaction effect by KK modes is 
not important in these models.

\section{Concluding remarks and discussions}   
In this paper, we have studied the excitement of
Kaluza-Klein (KK) modes in a higher $D$-dimensional
generalized Kaluza-Klein theory. 
We have considered two classes of models where
the extra dimensions are compactified on the 
sphere $S^d$ with $d=D-4$ and 
the direct product $S^{d_1}\times S^{d_2}$
with $d_1+d_2=D-4$.
Such compactifications give rise to a term which 
originates from the curvature of the internal space 
in the potential $U_1(\sigma)$ of a dilaton 
field $\sigma$. Since this potential does not have
a local minimum to stabilize the scale of the internal
space, we introduce the Casimir effect due to a
one-loop quantum correction. Then the $\sigma$ field 
oscillates as a massive scalar field 
around the local minimum at
$\sigma=0$, which corresponds to the present value
of the scale of compactifications.

Since the KK field acquires a mass
which can be expressed by the $\sigma$
field by compactifications on $S^d$ and 
$S^{d_1} \times S^{d_2}$, 
we can expect that KK modes will be
enhanced by parametric resonance.
The past work on this issue\cite{mukoh} is restricted in the
case of narrow resonance regimes where the resonance 
parameter $q$ is smaller than unity.
However, in the similar situation of preheating after
inflation with a quadratic potential, it is well known that 
resonance with $q \gg 1$ is much more efficient than
in the case of $q~\mbox{\raisebox{-1.ex}{$\stackrel
     {\textstyle<}{\textstyle \sim}$}}~1$. Hence we extend
past work on the excitation of KK modes 
to the case of $q \gg 1$ by making use of the stability
and instability chart of the Mathieu equation. 

In the case of $q~\mbox{\raisebox{-1.ex}{$\stackrel
     {\textstyle<}{\textstyle \sim}$}}~1$, 
the first resonance does not occur for 
both compactifications by $S^d$ and $S^{d_1}\times S^{d_2}$.
Although there are some situations where the KK
field is located in the second, third, $\cdots$ 
instability bands at the beginning of the coherent
oscillation of the $\sigma$ field,
parametric resonance soon becomes ineffective with the 
decrease of $q$ due to the expansion of the universe.
In this case, since the creation rate of KK modes can not surpass 
the Hubble expansion rate, the 
enhancement of KK modes is inefficient.

Even in the case of $q \gg 1$, we have found that 
the growth of KK modes does not take place
both by analytic approaches and numerical integrations.
The $\sigma$ field begins 
to oscillate coherently when the amplitude of the $\sigma$ field
drops down to $\epsilon \approx 0.4$,
where $\epsilon$ is defined by $(\ref{C40})$.
Since the relation of resonance parameters $A_k$ and
$q$ are expressed as $A_k=2q/\{(d+2)\tilde
{\sigma}(t)\}+4(k/m_{\sigma}a)^2$
for the $S^d$ case, and 
$A_k=2q/\{(d_1+d_2+2)\tilde
{\sigma}(t)\}+4(k/m_{\sigma}a)^2$
for the $S^{d_1} \times S^{d_2}$ case,
the KK field exists
in the region of 
$A_k~\mbox{\raisebox{-1.ex}{$\stackrel
     {\textstyle>}{\textstyle\sim}$}}~5q$ where 
instability bands are few at the beginning.
The amplitude of $\sigma$ decreases with the 
passage of time, and the KK field evolves in the 
regions where instability bands are further few.
As a result, KK modes are not relevantly enhanced 
even for the case of $q \gg 1$.
We have numerically confirmed this fact,
and found that the excitation of KK modes is
inefficient in any parameters in two classes of models
of compactifications.

Since we find that KK modes are not 
overproduced by parametric resonance by  
compactifications of $S^d$ and 
$S^{d_1}\times S^{d_2}$, this kind of 
compactification may not be ruled out from a 
cosmological point of view, because the energy density
of KK modes will not overclose the universe
in the radiation dominant era.
However, at the stage of preheating after inflation, 
scalar fields coupled to an inflaton field can be strongly
enhanced by parametric resonance\cite{KLS}.
If the KK field is coupled to inflaton, the 
enhancement of KK modes would also occur
in the preheating stage.
Recently, Mazumdar and Mendes\cite{MM} 
considered the excitement of the dilaton field $\sigma$
as well as the Brans-Dicke field during the preheating phase
in generalized Einstein theories.
Since they compactified the extra dimensions on torus
which does not have the curvature of the internal space,
the potential of dilaton $U_1(\sigma)$
does not appear in their model.
Taking into account the growth of metric perturbations
during preheating\cite{MET}, it was found that the dilaton field $\sigma$
can be effectively enhanced even when dilaton does not 
couple to inflaton. Although they did not consider 
the enhancement of KK modes, 
there will be a possibility that KK modes are 
strongly enhanced in the preheating phase even in the case
where the KK field does not directly couple to
inflaton by the growth of metric perturbations.
In this model, back reaction effects would play 
an important role for the termination of resonance.
Although it is technically difficult to deal with 
back reaction issues including
second order metric perturbations in a consistent way, 
it is of interest how KK modes are enhanced 
in the preheating phase.
These issues are under consideration.

\section*{ACKOWLEDGEMENTS}
The author would like to thank Kei-ichi Maeda,
Takashi Torii, Kunihito Uzawa, and Hiroki Yajima 
for useful discussions. 
This work was supported partially by a Grant-in-Aid for  Scientific
Research Fund of the Ministry of Education, Science and Culture
(No. 09410217), and by the Waseda University 
Grant for Special Research Projects.
Recently, Uzawa, Morisawa, and Mukohyama\cite{UMM} considered the 
excitement of Kaluza Klein modes including metric perturbations
in the narrow resonance case, and found that quanta of these modes
are not enhanced sufficiently. This result is consistent with our results
obtained in this paper. 

\newpage

\newpage
\begin{flushleft}
{ Figure Captions}
\end{flushleft}
\noindent
\parbox[t]{2cm}{FIG. 1:\\~}\ \
\parbox[t]{12cm}
{The potential $U_1(\sigma)$ which is obtained by 
introducing the Casimir effect in the $S^d$ compactification
with $d=2$. The potential has a minimum at $\sigma=0$
and a local maximum at $\sigma_*/\sigma_0=0.50$.
If we choose larger values of $d$, $\sigma_*/\sigma_0$
becomes smaller.
The shape of this potential 
in the case of the $S^{d_1} \times S^{d_2}$ compactification
is the same as in the case of the $S^d$ compactification.
}\\[1em]
\noindent
\parbox[t]{2cm}{FIG. 2:\\~}\ \
\parbox[t]{12cm}
{The schematic diagram of the Mathieu chart.
The lined regions denote the instability bands.
There exists narrow instability bands around $A_k=j^2$
and $q<1$ with positive integer $j$.
Although there are many instability bands
in the regions of $q \gg 1$, 
they are few for $A_k \ge 5q$.
}\\[1em]
\noindent
\parbox[t]{2cm}{FIG. 3:\\~}\ \
\parbox[t]{12cm}
{The evolution of $\sigma$ as a function of $t$
in the case of the $S^d$ compactification with $d=2$, $l=1$
and $k=0$.
We choose the initial value of $\sigma$ as 
$\sigma/\sigma_0=0.4$. After the first stage of rolling down,
the $\sigma$ field begins to oscillate coherently as Eq.~(3.8).
The dimensionless amplitude $\tilde{\sigma}(t)$
in Eq.~(3.8) decreases as $\tilde{\sigma}(t) \sim
1/t$ with the initial value of
$\tilde{\sigma}(t_{\rm co}) \approx 0.1$.
}\\[1em]
\noindent
\parbox[t]{2cm}{FIG. 4:\\~}\ \
\parbox[t]{12cm}
{The evolution of the real part of the Kaluza-Klein 
mode $\phi_k$ as a function of $t$ in the case of the 
$S^d$ compactification with $d=2$, $l=1$, and $k=0$.
Although the $\phi_k$ field is initially in the second 
instability band with $q~\mbox{\raisebox{-1.ex}{$\stackrel
     {\textstyle<}{\textstyle \sim}$}}~1$, the expansion of the
universe makes parametric resonance ineffective.
As a result, the growth of $\phi_k$ can not be expected.
}\\[1em]
\noindent
\parbox[t]{2cm}{FIG. 5:\\~}\ \
\parbox[t]{12cm}
{The evolution of the real part of the Kaluza-Klein 
mode $\phi_k$ as a function of $t$ in the case of the 
$S^d$ compactification with $d=2$, $l=100$, and $k=0$.
Although the initial value of $q$ is large as $q \gg 1$,
the $\phi_k$ field moves in the regions of
$A_k \ge 2q/\{(d+2)\tilde{\sigma}(t)\}$ where instability
bands are few. Hence parametric resonance is inefficient.
}\\[1em]
\noindent


\begin{thebibliography}{99}
\bibitem{comment}
Although the case of $B \le 1$ is considered in 
Ref.~\cite{mukoh}, we can easily verify that 
this does not hold for positive integers 
$l_1, l_2$ and $d_1, d_2$.
\end{thebibliography}
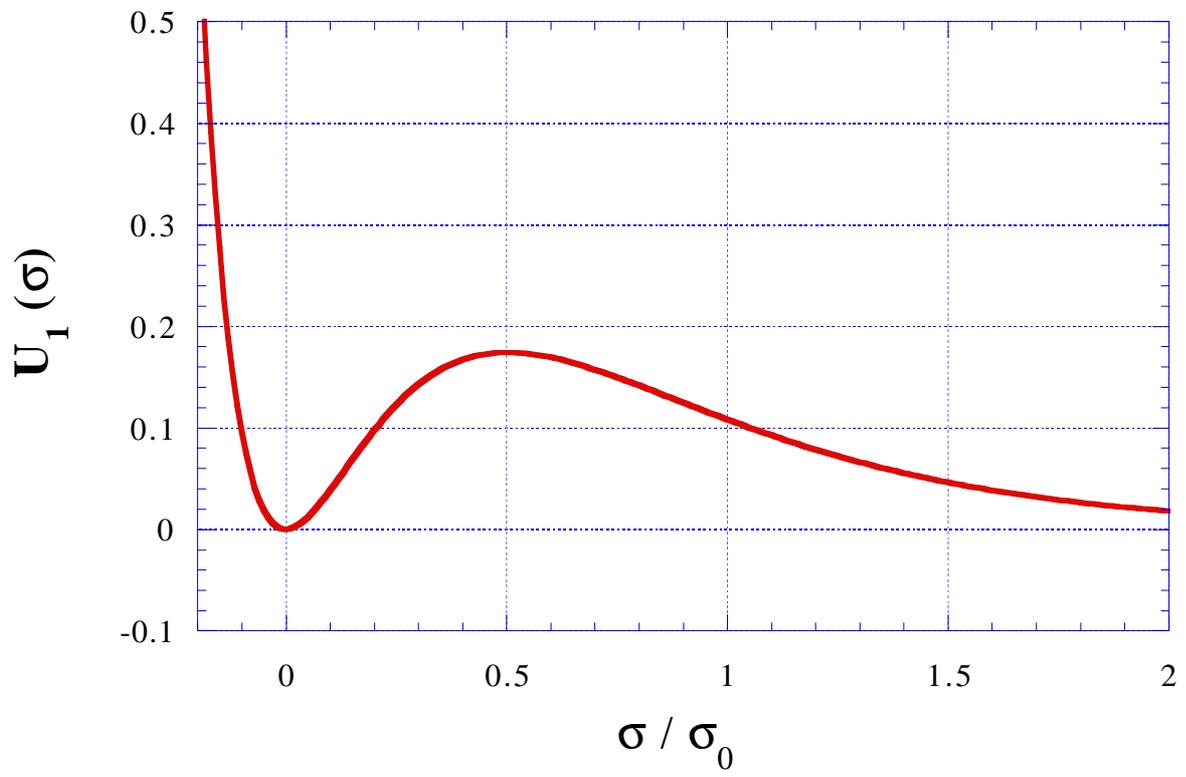
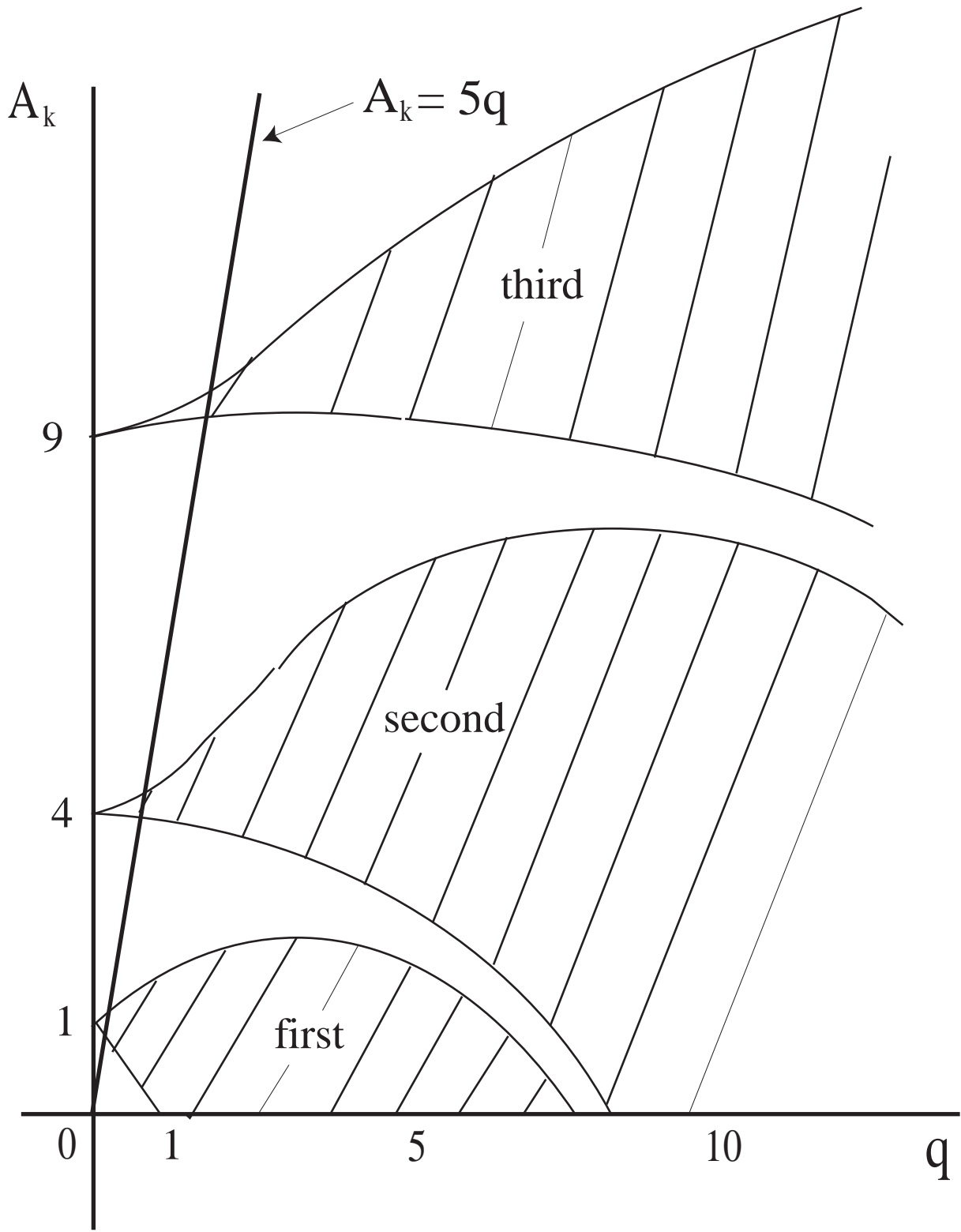
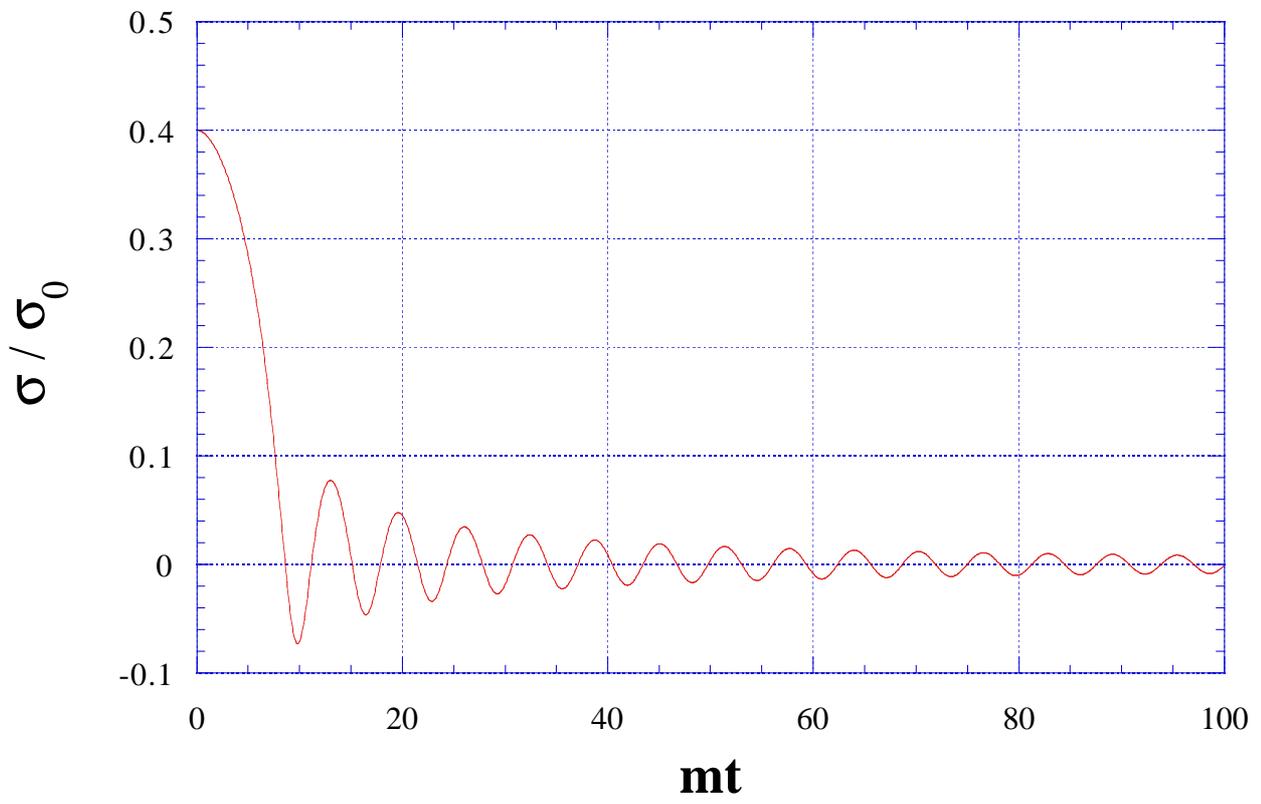
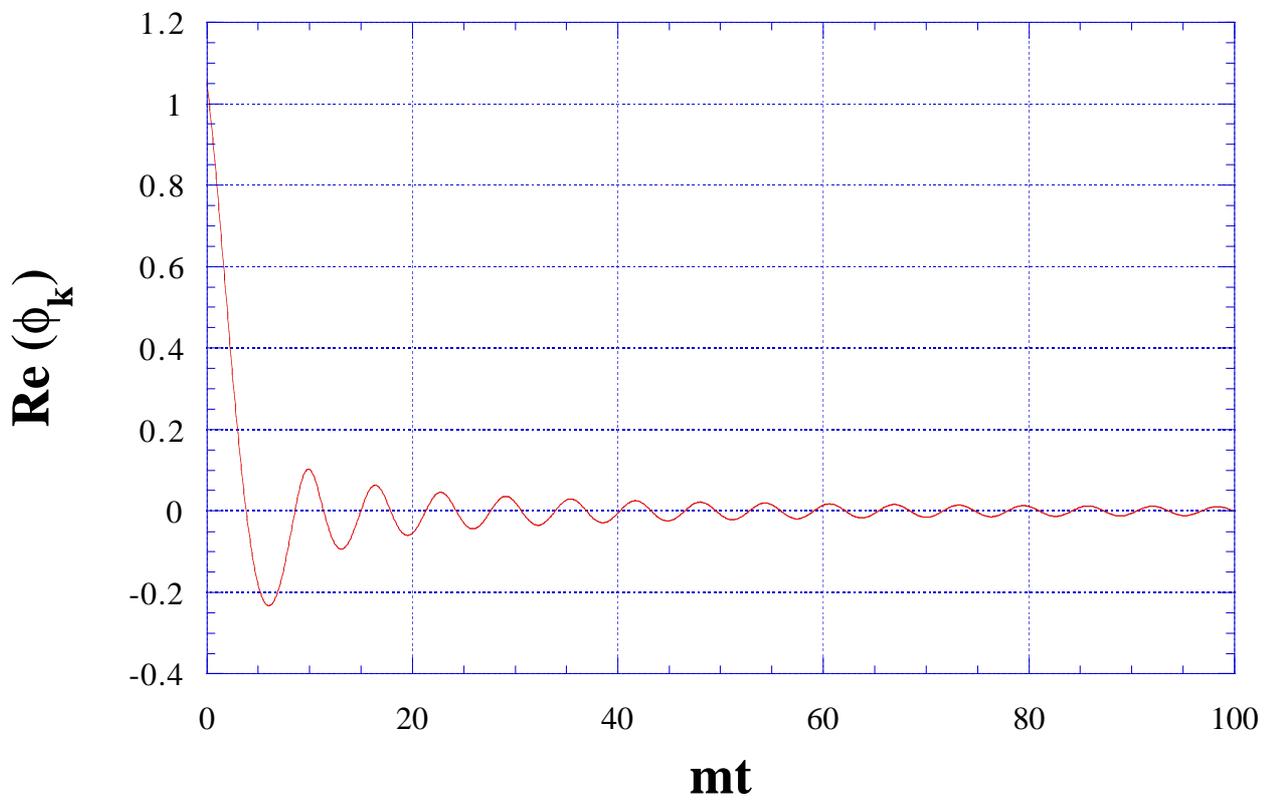
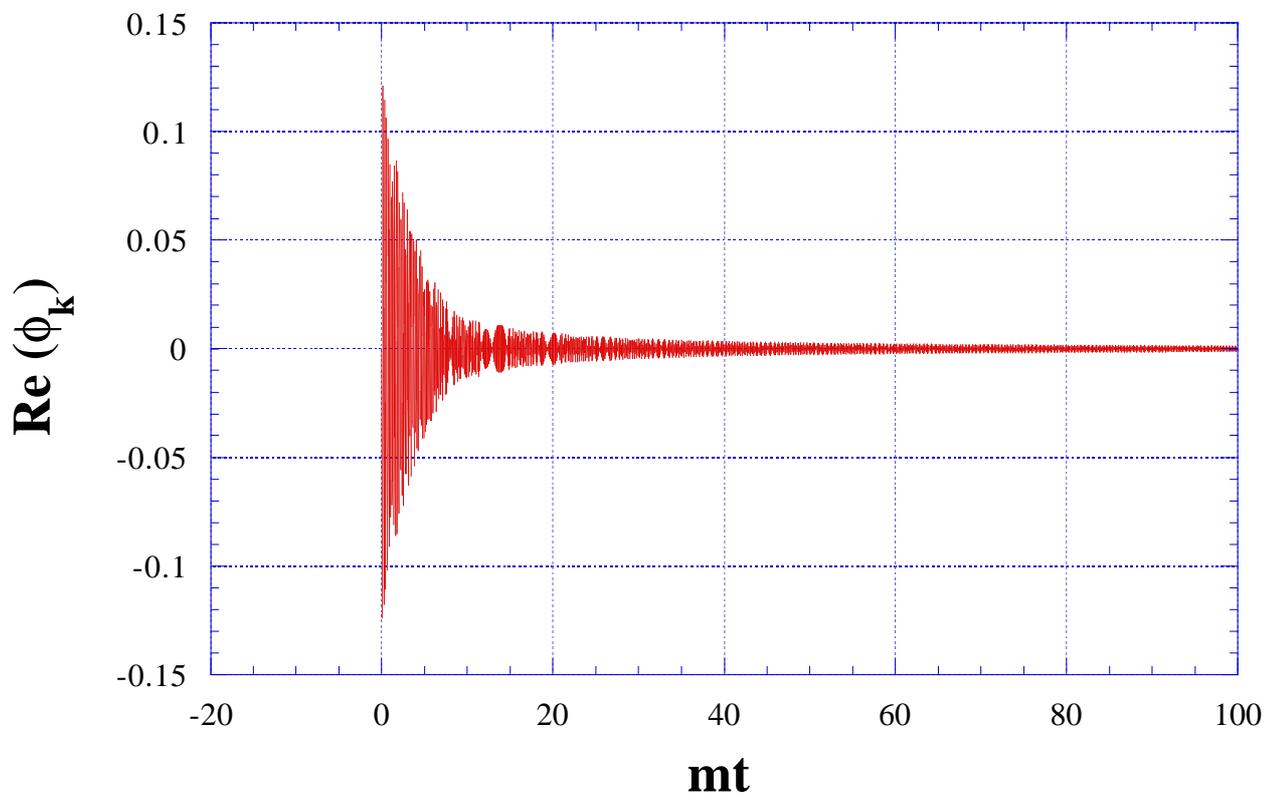
\end{document}